\documentclass[12pt,a4paper,twoside,english]{article}
\usepackage[T1]{fontenc}
\usepackage[latin1]{inputenc}
\usepackage{amssymb}
\usepackage{graphicx}



\usepackage{babel}
\begin{document}

\title{Phase diagram in Quantum Chromodynamics}

\date{{}}

\author{{\normalsize{M. Apostol }}\\
{\normalsize{Department of Theoretical Physics, Institute of Atomic
Physics, }}\\
{\normalsize{Magurele-Bucharest MG-6, POBox MG-35, Romania }}\\
{\normalsize{email: apoma@theory.nipne.ro}}}
\maketitle
\begin{abstract}
It is suggested that the hadronization of the quark-gluon plasma is
a first-order phase transition described by a critical curve in the
temperature-(quark) density plane which terminates in a critical point.
Such a critical curve is derived from the van der Waals equation and
its parameters are estimated by using the theoretical approach given
in M. Apostol, Roum. Reps. Phys. \textbf{59} 249 (2007); Mod. Phys.
Lett. \textbf{B21} 893 (2007). The main assumption is that quark-gluon
plasma created by high-energy nucleus-nucleus collisions is a gas
of ultrarelativistic quarks in equilibrium with gluons (vanishing
chemical potential, indefinite number of quarks). This plasma expands,
gets cool and dilute and hadronizes at a certain transition temperature
and transition density. The transition density is very close to the
saturation density of the nuclear matter and, it is suggested that
both these points are very close to the critical point $n\simeq1fm^{-3}$
(quark density) and $T\simeq200MeV$ (temperature). 
\end{abstract}


As it is well known, the Quantum Chromodynamics (QCD) developed in
the past 50 years describes the quark-gluon strong interaction. The
QCD lagragian is 
\begin{equation}
L=-\frac{1}{4}G_{\mu\nu}^{a}G_{a}^{\mu\nu}+\overline{\psi}_{f\alpha}\left[i\gamma^{\mu}(\partial_{\mu}\delta^{\alpha\beta}+igt_{a}^{\alpha\beta}A_{\mu}^{a})-m_{f}\delta^{\alpha\beta}\right]\psi_{f\beta}\,\,\,,\label{1}
\end{equation}
 where 
\begin{equation}
G_{\mu\nu}^{a}=\partial_{\mu}A_{\nu}^{a}-\partial_{\nu}A_{\mu}^{a}+gf_{bc}^{a}A_{\mu}^{b}A_{\nu}^{c}\label{2}
\end{equation}
 and 
\begin{equation}
\left[t_{a},t_{b}\right]=if_{ab}^{c}t_{c}\,\,.\label{3}
\end{equation}
$A_{\mu}^{a}$ are gluon potentials, with $\mu$ the Lorentz label
and $a=1...8$ the gluon label; $G_{\mu\nu}^{a}$ are the gluon fields,
$g$ is the coupling constant and $f_{bc}^{a}$ is the structure factor
of the SU(3) group. The eight $3\times3$ matrices $t_{a}$ ($\alpha,\beta=1,2,3$)
are the SU(3) generators. $\psi_{f\alpha}$ are the quark fields (bispinors),
labelled by flavour $f=1...6$ and color $\alpha=1,2,3$, with mass
$m_{f}$; $\gamma^{\mu}$ are the Dirac matrices. (For a review of
QCD the reader can consult the recent Refs. \cite{key-1}-\cite{key-3}).

The lagrangian given by equation (\ref{1}) is constructed by close
analogy with the Quantum Electrodynamics (QED), with two major differences:
the Yang-Mills fields (quadratic term in the gluon fields in equation
(\ref{2})) and the underlying SU(3) symmetry (color group). Perturbation-theory
calculations indicate that effective (renormalized) coupling strength
becomes weak for high-energy processes (and short distances), a phenomenon
known as quark de-confinement; the quarks and gluons are free at high
energy and get confined at low energy; for instance, they are bound
in hadrons (mesons and baryons), which are color-singlet states (zero
color charge).\cite{key-4,key-5} The confinement phenomenon, which
is opposite to QED, is due to the non-linear Yang-Mills contribution.
The usual type of calculations in QCD is the lattice-gauge theory
calculations,\cite{key-6} the results being often of a more qualitative
nature. Various simplifications are customary in calculations, in
particular the limitation to only (the lightest) $u$ and $d$ quarks,
whose mass is set equal to zero. In this case, the chiral symmetry
(handedness) of the theory should be broken at low energies. Broken
symmetries, associated phase transitions and, in general, methods
borrowed from condensed matter physics, are employed with the hope
of getting more quantitative results in QCD.\cite{key-2} 

In nucleus-nucleus collisions high energy can be transferred to the
internal structure of the nucleons (\emph{e.g.}, $1TeV$ per nucleon
as compared with the nucleon binding energy $1GeV$), such that we
may expect the liberation of quarks and gluons for a short while,
followed by a quick hadronization.\cite{key-7}-\cite{key-11} An
ultrarelativistic gas of quarks can be formed in such collisions,
reaching quickly the thermal equilibrium at a temperature produced
by the collision energy; we may speak of a quark-gluon plasma, with
a threshold (ignition) temperature (of cca $125-180MeV$), and a hadron-quark-gluon
plasma transition.\cite{key-12}-\cite{key-14} In a high-energy collision
the plasma expands, its volume and number of quarks and gluons (at
local equilibrium) increase, the quark density and temperature decrease,
and the quarks in the outer shell hadronize.\cite{key-15} It is tempting
to assign a second kind to such a transition, corresponding to the
broken chiral symmetry with zero-mass quarks, but the real situation
involves non-vanishing masses. We may only assume that the hadron-quark-gluon
plasma is first order, involving a hadron binding energy, very similar
with the van der Waals liquid (solid)-gas transition.\cite{key-16,key-17}
As it is well known, the van der Waals isotherms are given by 
\begin{equation}
(p+an^{2})(1-bn)=nT\,\,\,,\label{4}
\end{equation}
 where $p$ is the presssure, $n$ is the density, $T$ is the temperature
and $a,\, b$ are constants. Since the van der Waals characteristic
pressure is $\sim an^{2}$, we may also assume $p=cn^{2}$ at transition,
where $c$ is a constant. The van de Walls isotherms become 
\begin{equation}
(a+c)n(1-bn)=T\,\,\,,\label{5}
\end{equation}
 or 
\begin{equation}
T=-Bn^{2}+An\,\,\,,\label{6}
\end{equation}
 where $A$ and $B>0$ are constants . We can see that equation (\ref{6})
is equivalent with the van der Waals equation (\ref{4}) for zero
pressure. Since $\partial T/\partial n<0$ for a physical transition,
we can see that we should consider the above equation from $n=A/2B$
up to $n=A/B$ ($T>0$), so we should have $A>0$ (descending branch
of the second-order trinomial in equation (\ref{6})). Under these
conditions the above equation (\ref{6}) gives the curve corresponding
to the hadron-quark-gluon plasma transition in the $(n,\, T)$ plane,
the point $n_{c}=A/2B$, $T_{c}=A^{2}/4B$ being the critical point.
Equation (\ref{6}) can also be written as $T=An[1-(B/A)n]$, where
we can see that the ratio $B/A$ is a limiting volume, which may be
viewed as corresponding to a nominal \textquotedbl{}volume\textquotedbl{}
$v_{n}$ of the quarks, $B/A=v_{n}$; the quark density $n$ can be
written as $n=N/V=1/v_{q}$, where $v_{q}$ is the mean volume assigned
to a quark in the volume $V$ occupied by $N$ quarks. For a mixture
of various quark species, the density $n$ can be generalized to the
mean density involving partial densities. 

Now we describe briefly the theoretical approach given in Ref. \cite{key-15},
because it gives us access to the parameters $A$ and $B$ in equation
(\ref{6}). We consider a nucleus with $N_{n}$ nucleons in a volume
$V_{0}=R_{0}^{3}$, where $R_{0}$ is, approximately, the radius of
the nucleus (for simplicity, we leave aside the numerical factor $4\pi/3$);
the nucleus is subjected to a high-energy collision, with an energy
(per nucleon) $E/N_{n}=1TeV$, for instance. We limit ourselves to
the lightest quarks $u$ and d, for which we may neglect their mass
at these energy values ($m_{u}\simeq2MeV$, $m_{d}\simeq5MeV$). In
general, the energy is dominated by gluons, except for assuming an
ultrarelativistic gas of an indefinite number of quarks (vanishing
chemical potential) in equilibrium with gluons, \emph{i.e.} a quark-gluon
plasma. In this case, the plasma energy (quarks plus gluons, almost
equal energy) is given by 
\begin{equation}
E=VT^{4}/(\hbar c)^{3}\label{7}
\end{equation}
 and the mean number of quarks is 
\begin{equation}
N=VT^{3}/(\hbar c)^{3}\label{8}
\end{equation}
 (up to some immaterial numerical factors). (The pressure is $p=E/3$,
the entropy is $S\simeq4E/3T\simeq N$ and the density is given by
$T=\hbar cn^{1/3}$; $\hbar$ is Planck's constant ad $c$ is the
speed of light). At the initial moment we have $E_{0}=V_{0}T_{0}^{4}/(\hbar c)^{3}$
and $N_{0}=V_{0}T_{0}^{3}/(\hbar c)^{3}$; for an energy $E/N_{n}=1TeV$
we get $T_{0}=1GeV$ and $N_{0}=10^{3}N_{n}$ ($N_{n}\simeq100$),
assuming a nucleon radius $a=2fm$ and $R_{0}=aN_{n}^{1/3}$ ($\hbar c=200MeV\cdot fm$).
This plasma expands in time according to the laws%
\footnote{Compare with the hydrodynamical model of particle production, Refs.
\cite{key-18}-\cite{key-21}.%
}
\begin{equation}
\begin{array}{c}
R=R_{0}(1+ct/R_{0})\,\,,\,\, V=V_{0}(1+ct/R_{0})^{3}\,\,,\\
\\
T=T_{0}(1+ct/R_{0})^{-3/4}\,\,,\,\, N=N_{0}(1+ct/R_{0})^{3/4}\,\,;
\end{array}\label{9}
\end{equation}
 its density goes like 
\begin{equation}
n=N/V=n_{0}(1+ct/R_{0})^{-9/4}=n_{0}(T/T_{0})^{3}\label{10}
\end{equation}
 or, using $N_{0}=V_{0}T_{0}^{3}/(\hbar c)^{3}$ (\emph{i.e.} $n_{0}=(T_{0}/\hbar c)^{3}$),
\begin{equation}
T=\hbar cn^{1/3}\,\,;\label{11}
\end{equation}
if we put here the quark density in the cold nucleus ($n\simeq N_{n}/V_{0}$,
or $n\simeq3N_{n}/V_{0}$) we get the threshold (ignition) temperature
$100-150MeV$ (for $a=2fm$; the values $125-180MeV$ given above
are obtained for $a=1.5fm$). We can see that, during expansion, plasma
gets cool and the quark density decreases according to equations (\ref{9});
at the same time, the energy is conserved and the entropy increases.
Equation (\ref{11}) defines also the chemical potential of a degenerate
ultrarelativistic gas of quarks ($\mu=\hbar cn^{1/3}$). 

Further on, a mechanism of condensation (hadronization) has been put
forward in Refs. \cite{key-15,key-22} (a first-order phase transition).
The transition temperature is given by
\begin{equation}
T_{t}\simeq T_{q}(T_{q}/T_{m})^{1/2}\,\,\,,\label{12}
\end{equation}
where $T_{q}$ is a characteristic quark temperature and $T_{m}\simeq m_{0}c^{2}$
is a characteristic temperature given by the average mass $m_{0}$
of the condensed quarks (up to some immaterial numerical factors).
It is shown in Ref. \cite{key-15} that only a fraction $f$ of the
quark number is affected by hadronization (and dominates the hadronization
process, with a classical Boltzmann statistics), so that we have in
fact 
\begin{equation}
T_{t}\simeq f^{1/2}T_{q}(T_{q}/T_{m})^{1/2}\,\,.\label{13}
\end{equation}
At transition $T_{t}=T_{q}$ and 
\begin{equation}
T_{t}=f^{-1}T_{m}\,\,,\:\hbar cn_{t}^{1/3}\simeq f^{-1}m_{0}c^{2}\,\,\,,\label{14}
\end{equation}
an expected and plausible result. We can see that the transition density
\begin{equation}
n_{t}=f^{-3}\left(\frac{m_{0}c}{\hbar}\right)^{3}\label{15}
\end{equation}
 is related to the Compton wavelength $\hbar/m_{0}c$ of the \textquotedbl{}average\textquotedbl{}
condensed quarks. We can take $m_{0}=4MeV$ (mean mass of the $u$
and $d$ quarks), and get $n_{t}=f^{-3}(50fm)^{-3}$. In Ref. \cite{key-15}
it is suggested that fraction $f$ is given approximately by $f=1/N_{0}^{1/3}$$\simeq2\times10^{-2}$
(an argument derived from the saturation of the nuclear forces), so
we get the transition density $n_{t}\simeq1fm^{-3}$. It corresponds
to a transition temperature $T_{t}=f^{-1}T_{m}\simeq200MeV$ (slightly
above the ignition threshold) and a transition radius $R_{t}=R_{0}(n_{t}/n_{0})^{-4/9}=R_{0}(T_{0}/T_{t})^{4/3}\simeq8R_{0}$.
This hadronization happens after $t=5\times10^{-23}s$ from the collision
and involves $fN_{t}=fN_{0}(R_{t}/R_{0})^{3/4}\simeq100N_{n}$ hadronized
quarks. (The transition implies a latent heat, discontinuities of
the thermodynamic potentials, etc, as for a first-order (van der Waals)
transition). We can see that the transition density $n_{t}=1fm^{-3}$
is very close to the saturation density of the nuclear matter. 
\begin{figure}
\noindent \begin{centering}
\includegraphics[clip]{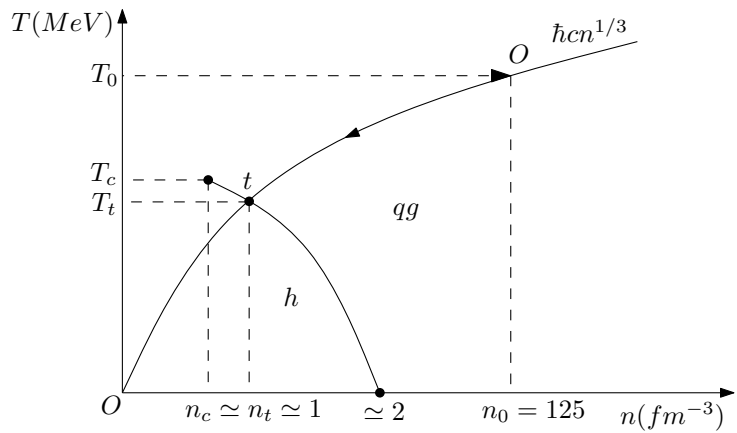}
\par\end{centering}

\caption{Hadronization of the quark-gluon plasma. Phase diagram temperature
($T$) \emph{vs} quark density ($n$). Note the hadronization curve
$T=\hbar cn^{1/3}$.}

\end{figure}

The transition temperature and density must obey the critical curve
give by equation (\ref{6}). With temperature measured in $MeV$ and
density measured in $fm^{-3}$ we get 
\begin{equation}
200=A-B\,\,\,,\label{16}
\end{equation}
 which gives a relation between the two parameters $A$ and $B$ of
the critical curve. Since the transition density is very close to
the saturation density of the nuclear matter, we may take tentatively
the critical density $n_{c}=A/2B=n_{t}=1fm^{-3}$; we get $A=400$
and $B=200$ (and the critical temperature $T_{c}=A^{2}/4B=T_{t}=200MeV$);
the density where the critical curve crosses the $n$-axis is $A/B=2fm^{-3}$.
It is worth noting that the critical point $n_{c}=1fm^{-3}$, $T_{c}=200MeV$
derived here can have a universal character; indeed, it lies on the
curve $T=\hbar cn^{1/3}$ for $n$ equal to the saturation density
of the nuclear matter (\emph{e.g.}, $n=1fm^{-3}$).

The hadronization of the quark-gluon plasma is shown in Fig. 1. According
to the description given above the hadronization process starts with
the creation of a quark-gluon plasma at the initial quark density
$n_{0}$ (e.g., $\simeq10^{3}N_{n}/V_{0}=125fm^{-3}$) and temperature
$T_{0}$ (\emph{e.g.}, $1GeV$), followed by a cooling of the plasma
along the curve $T=\hbar cn^{1/3}$ untill it encounters the critical
curve at the transition point $n_{t}$ (\emph{e.g.}, $1fm^{-3}$)
and $T_{t}$ (\emph{e.g.}, $200MeV$) where the hadronization occurs.
With our units ($MeV$ and $fm$) the curve $T=\hbar cn^{1/3}$ reads
$T=200n^{1/3}$. The transition temperature is very close to the saturation
density of the nuclear matter and, very likely, it is very close to
the critical density.\cite{key-23} 

\textbf{Acknowledgments.} The author is indebted to MPD-NICA JINR
Dubna (Russia) Project, especially to A. G. Litvinenko and I. Cruceru,
for many useful discussions, and to the members of the Seminar of
the Laboratory of High Energy Physics, JINR Dubna, for a thorough
analysis of this work.

\end{document}